\documentclass[11pt]{article}

\title {Report of the HPC Correctness Summit \\
   Jan 25--26, 2017, Washington, DC}

\usepackage[colorlinks]{hyperref}
\usepackage[colorinlistoftodos]{todonotes}
\usepackage{paralist}
\usepackage{verbatim}
\usepackage{xcolor}
\usepackage{adjustbox}
\usepackage{graphicx}
\usepackage{booktabs}
\usepackage{xparse}
\usepackage{amssymb}
\usepackage{color, colortbl}
\usepackage{wrapfig}
\usepackage{tcolorbox}

\hypersetup{
     colorlinks   = true,
     citecolor    = blue,
     linkcolor = blue,
     linkbordercolor= blue,
     urlcolor = blue
}

\NewDocumentCommand{\rot}{O{45} O{1em} m}{\makebox[#2][l]{\rotatebox{#1}{#3}}}%

\newenvironment{WrapText}[1][r]
{\wrapfigure{#1}{0.5\textwidth}\tcolorbox}
{\endtcolorbox\endwrapfigure}

\newenvironment{WrapTextLeft}[1][l]
{\wrapfigure{#1}{0.5\textwidth}\tcolorbox}
{\endtcolorbox\endwrapfigure}
\newcommand\ggc[1]{{{#1}}}

\newcommand\ggcmt[1]{\todo[inline, size=\small, color=green!40]{GG: #1}}
\newcommand\ggcmtside[1]{\todo[size=\small, color=green!40]{GG: #1}}
\newcommand\phcmt[1]{\todo[inline, size=\small, color=orange!40]{PH: #1}}
\newcommand\cicmt[1]{\todo[inline, size=\small, color=red!40]{CI: #1}}
\newcommand\skcmt[1]{\todo[inline, size=\small, color=yellow!35]{SK: #1}}
\newcommand\ilcmt[1]{\todo[inline, size=\small, color=blue!40]{IL: #1}}
\newcommand\rlcmt[1]{\todo[inline, size=\small, color=brown!40]{RL: #1}}
\newcommand\kscmt[1]{\todo[inline, size=\small, color=blue!20]{KS: #1}}
\newcommand\sscmt[1]{\todo[inline, size=\small, color=yellow!60]{SS: #1}}
\newcommand\ascmt[1]{\todo[inline, size=\small, color=orange!60]{AS: #1}}
\newcommand\ignore[1]{}
\newcommand\ASGNMT[1]{}

\newcommand{\lb}{\texttt{\char`\{}}
\newcommand{\rb}{\texttt{\char`\}}}

\begin{document}

\ignore{
\ggcmt{hi}
\phcmt{hi}
\cicmt{hi}
\skcmt{hi}
\ilcmt{hi}
\rlcmt{hi}
\kscmt{hi}
\sscmt{hi}
\ascmt{hi}
}

\author{
  {\sc report authors} 
  \ \\
  \ \\
  \begin{tabular}{lr}
Ganesh Gopalakrishnan & University of Utah\\
Paul D.\ Hovland & Argonne National Laboratory\\
Costin Iancu & Lawrence Berkeley National Laboratory\\
Sriram Krishnamoorthy & Pacific Northwest National Laboratory\\
Ignacio Laguna & Lawrence Livermore National Laboratory\\
Richard A.\ Lethin & Reservoir Labs, Inc., Yale University\\
Koushik Sen & University of California, Berkeley\\
Stephen F.\ Siegel & University of Delaware\\
Armando Solar-Lezama & Massachusetts Institute of Technology\\
 \end{tabular}
}
\date{May 21, 2017}

\maketitle

\vspace{-3ex}
\begin{center}
{\sc DISCLAIMER}
\end{center}

This {\bf preliminary version of our report} was prepared as an account of a summit sponsored by the U.S. Department of Energy. Neither the United States Government nor any agency thereof, nor any of their employees or officers, makes any warranty, express or implied, or assumes any legal liability or responsibility for the accuracy, completeness, or usefulness of any information, apparatus, product, or process disclosed, or represents that its use would not infringe privately owned rights. Reference herein to any specific commercial product, process, or service by trade name, trademark, manufacturer, or otherwise, does not necessarily constitute or imply its endorsement, recommendation, or favoring by the United States Government or any agency thereof. The views and opinions of document authors expressed herein do not necessarily state or reflect those of the United States Government or any agency thereof. Copyrights to portions of this report (including graphics) are reserved by original copyright holders or their assignees, and are used by the Government’s license and by permission. Requests to use any images must be made to the provider identified in the image credits (if any) or the first author.

\clearpage
{
\hypersetup{linkbordercolor=blue}
\tableofcontents 
}

\clearpage

\section{Introduction\ASGNMT{Ganesh (lead), Richard}}
\label{sec:intro}
Technologies for verification and debugging have made significant strides in the context of general systems software. An investment in such technologies to make them applicable for High Performance Computing (HPC) could lead to substantial improvements in the productivity and sustainability of HPC software development. Such improvements will be essential to fully exploit new exascale computer architectures.
\ignore{------
The technology for verification and debugging in support of correctness has advanced to the point that there is a basis for confidence in success of investment in the application and advancement of such tools 
in High Performance Computing (HPC).  With such investment, the field of computational science will benefit from substantial improvements in the ability to exploit new exascale computer architectures, and from substantial improvements in productivity and sustainability of HPC software development.
-----}  Without such investment, there is the possibility of a substantial crisis in our ability to advance the field of HPC, as the complexity of our architectures, algorithms, and applications is moving beyond the ability of our developers.  As HPC is of strategic importance to our nation, forming the bedrock of its scientific and technological capabilities, such investment is highly warranted.

\subsection{Reasons for the correctness crisis}

\begin{figure}
\centering
\begin{adjustbox}{width=1.15\textwidth,center}
\includegraphics[width=0.99\textwidth]{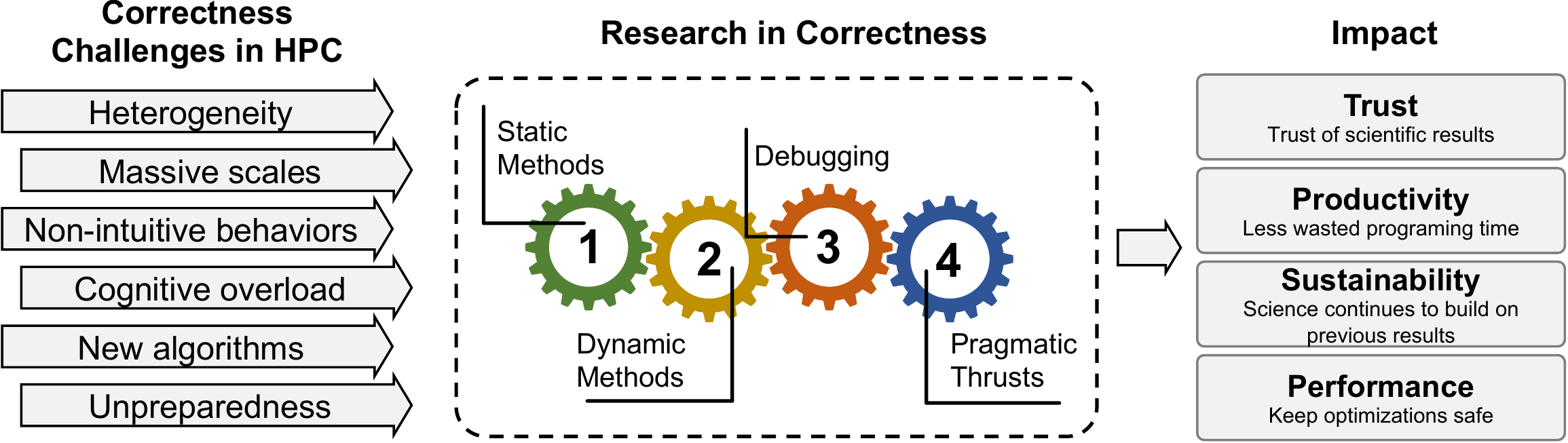}
\end{adjustbox}
\caption{Overview of the existing challenges in correctness for HPC and the research areas that need extensions to address these challenges.}
\label{fig:overview-challenges}
\end{figure}

While the general correctness problem in computer science is well
researched, specific reasons that cause HPC-specific correctness
methods to turn into an urgent priority include the following (see
Figure~\ref{fig:overview-challenges}
for an overview).

\paragraph{Growing heterogeneity:}

  Given the widening disparities between CPU, memory, and I/O speeds,
  computations will be supported by a heterogeneous architectures that include CPUs, GPUs, and special-purpose accelerators~\cite{hwu-heterogeneity}.
 Even today,
  programs in this space exist around a patchwork of semantic
  abstractions that are poorly understood individually, and
  whose emergent behavior is poorly understood.

\paragraph{Massive scale:}

  Attaining exascale will require proper coordination and synchronization
  across many tasks, threads, and processes~\cite{exascale-computing-project}.
  Disciplines that guide programming in this space do not exist, nor do testing methods that unearth defects in this space.
Left unchecked, this will lead to deployed systems that yield untrustable results or
crash during long-running simulations. 
Fixing the root cause of bugs in these
settings will incur huge latencies during which science
domain
experts may sit idle or be unproductively engaged with
debugging. These costs are known to be already very high (see the sidebars for some concrete examples of costly field bugs in HPC).
\ignore{----
but seldom highlighted in summary reports of HPC projects.
----}

\paragraph{Non-intuitive behaviors:}
 The push toward significant energy savings will lead to the use of
  reduced floating-point arithmetic, delayed state updates across weak
  memory consistency models, non-determinism caused by dynamic
  voltage/frequency scaling, fault recovery steps and inherent
  application non-determinism.  It is impossible to employ ad hoc
  debugging methods in these situations.

\paragraph{Cognitive overload:}
The manner in which people write as well as
configure full applications is
evolving in a direction where human reasoning about correctness is impractical.
  For example, the NWChemEx computational chemistry code is bringing together SPMD and task parallelism, multiple programming models and runtime interfaces, code generation, and dynamic and adaptive selection of execution configurations. 
In these settings, manually reasoning about correctness of the application, or even an individual execution, has gone beyond the scope of the individual developers;
the need to bring in more automated and/or
formal ways has become quite apparent.

\paragraph{New scalable algorithms:}
New algorithms for numerical computing offer the potential for major improvements in the asymptotic requirements for computation~\cite{ascac-2017}.  Kernel-independent and generalized Fast Multipole Methods (FMM) enable matrix vector multiply to be achieved in $O(N)$ time.  Support preconditioner methods will enable sparse systems of equations to be solved in near-linear time.  Randomized linear algebra and compressive methods will enable systems to be simulated in sampled or compressed form.  Such algorithms will bring complex new execution patterns and complex tradeoffs between precision, probability, and time.  The ability to exploit these algorithms will require tools to facilitate reasoning about the correctness of their application and implementation.

  %
\begin{WrapText}
\footnotesize
{\large \textbf{Undoing an optimization leads to a difficult bug}} \\

Tensor contraction expressions in Coupled Cluster methods involve products of multiple anti-symmetric tensors. In NWChem, these expressions are computed as a chain of pairwise tensor contraction to minimize operation count. An effort to reoptimize these contractions in the new generation of tensor contraction engine (TCE) by undoing the chaining and reoptimizing the contraction sequence in the existing TCE led to incorrect results. Isolating the source of the error was a manual process: transforming subsets of the contractions and checking for differences. After months of effort, it was found that, depending on the chaining, an additional coefficient needed to be introduced in a key intermediate step, referred to as symmetrization. Explicit specification of the optimization rules and checking whether or not the transformations satisfy the rules could have identified this bug quickly.
\end{WrapText}

\paragraph{Community unpreparedness:} \mbox{ }\\
  Compared to sequential programming abstractions that are more familiar
  to an application scientist, future HPC systems will involve a large
 slew of semantic abstractions (and relatively
 newer abstractions such as tasking)
 whose correct and efficient usage are
  not first nature to the broad application development community.
 
Ways to insulate application developers through
domain-specific languages (DSL) are badly needed; yet,
progress is lacking in this direction.
A related but severe problem is that
HPC application developers do not have the mindset
(or the necessary common repositories) for sharing
best practices (including
  sharing information on bugs and bug-fixes).
  This lack will hamper 
the transitioning of new verification research results into practice.

  \ignore{==> Ganesh addressed this above:
\todo[inline]{Should we include two more reasons: (1) use of DSLs, 
(2) use of task-based programing models?}
<==}

\subsection{What is in scope, what is not}


\paragraph{Current issues in scope:}
Any defect a programmer can correct by modifying and/or repairing
existing programs and/or their support runtime logic are well within the scope of this report.
These include
the classic sequential program bugs, errors relating to concurrency
(e.g., race conditions, incorrect programming under weak memory
models), and numerics (e.g., errors in realizing the numerical
algorithm using finite-precision floating-point numbers).  Defects that may
manifest only when a code is scaled up and not  during
lower-scale testing are also in scope.

We also consider defects that can be eliminated by disciplined code
 transformations from higher-level, or can be eliminated through
 better composition and software engineering practices.  Defect
 prevention methods that can be incorporated into best practices,
 pedagogy, mineable bug repositories, expert tutorials, and IDEs that
 can prevent or issue warnings about possible bugs are also of great
 importance, and are well within scope.

  \begin{WrapText}
\footnotesize
{\large \textbf{A Hard-To-Debug Large-Scale Error}}\\
A bug in ddcMD, a parallel molecular-dynamic code, manifested as an
intermittent hang when run at large scale on BlueGene/L at LLNL.
It took a significant amount of person hours and debugging effort 
to find the root cause: a message race in which a process could 
hang waiting for a message that was intercepted by 
another process. More specifically, the hang occurred when 
two independent instances of a user-level I/O layer
were simultaneously processing two separate sets of buffers---an infrequent pattern that occurred when a small data set 
was written immediately after a large data set. Due to the 
semantics of MPI send/recv operations and the use of fixed tags, 
messages from a small set could be confused for those for a 
large set and vice versa, thus triggering the hang. 
Later, after the bug was fixed by the programmer, the bug was 
used in a blind study, in which researchers developed a tool 
to isolate this class of bugs without having details of the error 
(more in~\cite{PACTLaguna:2012}). This shows that documenting 
bug cases can be useful in developing and testing advanced 
correctness tool.
\end{WrapText}


When defining the correctness of HPC programs, it is important to keep in mind that the behavior of a user-written program is heavily influenced by the
behaviors of the underlying libraries.
Not only are the sequential (e.g., numerical and C/Fortran) libraries important, the behavior of communication and runtime libraries (e.g., MPI and OpenMP) directly impact how a user program executes and whether it even makes forward progress.

In this context, it is important to detect
and eliminate
erroneous arguments supplied to library functions
that may cause a program crash. 
For example, MPI calls must adhere to conventions pertinent to
the source language (Fortran or C).
However, resource aspects of the runtime and communication libraries are a whole different matter.
For example, it is possible to write a user program that may be perfectly correct as far as the users' mental model of an ``idealized MPI library'' goes, but unfortunately a given MPI library may be unable to park all the asynchronous sends that the user program has issued.
Such user programs can either deadlock or crash an MPI library.

Most libraries are
underspecified, and their implementations often do not come with strong
guarantees, such as about the amount of resources provided (e.g., amount of
buffering) or whether forward progress or a response within a
deadline is guaranteed. 
These issues are clearly also important, but must be relegated to a longer-term pursuit that involves cooperation from library and runtime designers.

In the same vein,
the inability to control the evolving semantics of libraries and
programming languages must be kept in mind, requiring cooperation
among participant communities.   For instance, if a library
guarantees a certain order of accuracy for its results (e.g., ``9
digits of accuracy'') for specific platforms, we may not be prepared
to detect the violation of such contracts by another library
on a newer platform to which the code is ported.

\begin{WrapTextLeft}
\footnotesize
{\large \textbf{When More Than \textit{print} Debugging Is Needed}}\\
A scientist experienced hangs in a laser-plasma interaction code 
(named PF3D) when scaling it to 524,288 MPI processes on 
LLNL’s Sequoia BlueGene/Q system. The scientist spent months trying 
to debug the problem through print statements to no avail. 
Moreover, the scientist was unable to reproduce the hang at 
smaller scales where fully featured, heavyweight debuggers 
would be more plausible. Using STAT (Stack Trace Analysis Tool), 
the scientist was able to debug the problem, a race condition 
between two distinct but overlapping communication code regions. 
The bug was the result of the application migrating from one version to another more scalable but incompatible one. During migration, the application ran 
through a compatibility layer that introduced the race condition 
and ultimately caused the timing- and scale-dependent hangs 
(more in~\cite{CACM:Debugging}). This case shows that, although 
print debugging can be a useful debugging method, 
the HPC community can benefit from advanced correctness 
methods and tools to isolate bugs that otherwise can consume 
months of effort and millions of CPU hours to fix.

\end{WrapTextLeft}

\paragraph{Upcoming issues in scope during exascale:}
We also realize that
this report is being written in a timeframe where the underlying designs of exascale systems are experiencing significant disruptive changes.
In this era, hardware will often be poorly specified, particularly with regard to features related to
the memory model, concurrency and synchronization.  With exascale, new hardware features
for controlling voltage and frequency, are appearing, as well as advanced features for
task scheduling, communication, and synchronization. 
There will be heavy uses of heterogeneous types of memory (e.g., non-volatile, scratchpad spaces that do not provide cache coherence, etc.).
Consider the behavior
of an adaptive congestion algorithm~\cite{jiang2009indirect} in the communications fabric, which may 
affect or permute the ordering of delivery of messages. If such features are
not specified or are incorrectly specified to the runtime or MPI libraries, it will be impossible for any verification technology to guarantee correctness of the software running on it.

The performance behavior of hardware is also moving into the arena
of correctness and even safety, where exascale hardware systems are likely to be
over-provisioned with transistors, so that not all 
parts of the system can be used simultaneously while still remaining below
the power and cooling limits of the facility, and within the limits for safe and correct operation of the machine.  This will impose requirements
on firmware, system software and applications to maintain resource usage.  

All these
issues clearly point to an even greater 
demand for formal specifications from the hardware
vendors on these behaviors and requirements.
It will also correspondingly demand that our formal verification
and debugging tools use these hardware level formal specifications in order to provide overall correctness guarantees.

\begin{WrapText}
\footnotesize
{\large \textbf{Heterogeneity-caused arithmetic divergence results in deadlock}}\\
  In a recent project~\cite{xsede13-porting-bug},
an attempt to port some of the MPI processes to run on
Xeon-Phis while leaving others running on Xeons
caused a curious deadlock that took days to debug.
The root cause was the Xeon-Phis calculating the number of
messages to be sent (through an expression
$\lfloor p/c\rfloor $) differently from how the Xeons calculated
the number of messages to be received (also governed by $\lfloor p/c\rfloor$).
Unfortunately, the developers had not applied due precautions
to their compilation flags, resulting in $63$ messages 
being sent but only $62$ attempted to be received, which
then caused the deadlock.
This bug tells us that a 
combination of factors---processors
being different, floating-point 
roundoff differing
due to the inconsistent use of 
compiler optimization flags, and
the delicate MPI semantics allowing
the number of receives posted to exceed
the number of sends posted (but not
vice versa)---may  lead to bugs.
\end{WrapText}

\paragraph{The following issues are not directly within scope:} There are many issues that are important to keep in mind, but are best relegated to other pursuits that are better able to focus on them. 
We now mention a few examples of such issues (by no means exhaustive).
HPC programs may be brought down by
hardware logic errors in microprocessors, GPUs,
memory subsystems, and  buggy interconnect
protocol implementations.
Soft errors may corrupt program behavior, but are not considered human-introduced 
defects.
Version control and security-related issues are again somewhat tangential.
Finally, the numerical algorithm itself can be incorrect.
For instance,
errors in the design of the numerical scheme to
approximate the idealized mathematics,
including incorrectly scheduled coarse/fine meshing, lack of
conditioning of the problems, etc.,
can be considered {\em algorithmic}
defects and not software defects.

\subsection{Suggested research foci, targeted time-frames\ASGNMT{Ganesh (lead), Richard}}
%
%

 
 We now summarize some of the key short-term, medium-term, and long-term directions identified and elaborated in the rest of this report.

      \subsubsection{Short-term (1-2 years)}
      \noindent The following short-term foci are overdue, in order to bootstrap the process and bring the community together:

\begin{compactitem}
\item Launch efforts to apply existing best-of-breed tools to challenge problems, extend those tools, and generally work with HPC applications code as-is.
These tools include existing commercial tools as well as those being developed within the research community.

\item Advance these tools to address cross-cutting concerns adequately (e.g., tie-in to debuggers and formal tools, instrument existing OpenMP and MPI runtimes to produce event streams, standardize verification tool design around such event streams).

\item Bringing advances from the non-HPC community to HPC. These measures could begin as modest as ensuring the capture and sharing of bugs and their fixes, and in general, incorporating lessons from Empirical Software Engineering~\cite{menzies-empirical-swe-SI2-2017}.

\item Learn
from other communities. 
For example, 
study and adapt techniques for concurrency verification from the embedded system verification community. Also, adapt techniques for verifying numerical computations from the cyberphysical systems community.
\end{compactitem}

      \subsubsection{Medium-term (2-5 years)}
      \noindent The following medium-term directions deserve significant attention:

\begin{compactitem}
\item Correctness verification of important properties in common HPC software components,
including math libraries and widely-used runtime systems such as OpenMP and MPI.

\item Building infrastructure to document previously solved correctness issues in 
the form of bug databases, bug test cases, best testing practices, as well as lightweight 
mechanisms to automate the extraction of such cases in HPC centers.

\item Standardize interfaces to allow composability of correctness checkers, 
defect isolation tools, and debuggers.

\item Investments in the modeling and specification of numerical algorithms,
ontologies for the mathematics of the underlying algorithms.

\item Support for reasoning about statistical and randomized systems,
Uncertainty Quantification and Automatic Differentiation.

\end{compactitem}

      \subsubsection{Long-term (5 years and beyond)}
      \noindent Investments in these long-term directions will go a long way toward closing the gap between growing system complexity and verification capabilities:

\begin{compactitem}
\item A few ``moonshot'' projects, including  verification of fundamental logic and numerical properties in 
multi-physics applications.

\item Define   metrics 
for achievable and 
communicable levels of correctness, 
especially in simulations with geopolitical consequences, such as weather simulations.  
\end{compactitem}

\section{Rigorous Methods for Correctness\ASGNMT{Steve (lead), Ganesh, Richard, Sriram, Koushik, Paul}}
\label{sec:correctness-problem-hpc}


Formal methods are rigorous mathematical techniques for exhaustively checking that the
model of a system under analysis satisfies a set of desired properties.
The model in question could be: 
 (1)~a piece of code (e.g., the model of a numerical routine); or
\ignore{===
(2)~formalism underlying a particular
type of analysis (e.g., partial orders that describe a family of memory models, or a particular
formal model describing floating point arithmetic);
===}
(2)~a set of rules or axioms describing the behavior of some aspect of the system (for example, partial orders can describe the guarantees provided by a memory model, or a set of mathematical rules can describe the behavior of floating point arithmetic).
The second perspective is an example of formal methods   that are ``baked into'' analysis procedures or developed to
support specific lines of reasoning.

\ignore{===
or even 
(3)~formal models underlying
a compiler that produces a piece of stencil code based on its sketch.
 The latter two  
 ===}

\ignore{==>
At the very least, the extent of expert involvement must be contained (e.g., one particular numerical routine), with the correctness of all code variants of that routine implied by formalized transformations.

 The main reason to emphasize formal methods is that HPC can learn
from the hardware industry where
{\em almost all chips that matter} are subject to some extent of formal analysis.
Often {\em semi-formal} adaptations work best in practice where the driving theory/formalism is the same, but the checking steps are bounded in some sense (e.g., concretizing some inputs, bounding the depth of exploration, selecting critical subsystems for analysis while under-approximating their environment, etc.).
Here are two examples:
\begin{compactitem}
\item No microprocessor gets shipped without its cache coherence protocols being 
subject to formal analysis and correction; this trend started with pioneering efforts
such as Loewenstein's
adaptation of David Dill's Murphi tool around
1991 at Sun Microsystems.
Even today, Murphi is the language of choice for modeling and verifying industrial cache coherence protocols.

<==}

Following the well-known Intel Pentium fiasco, all major chip manufacturers 
 have now adopted formal analysis to
 verify floating-point hardware.
 In a recent project at Intel~\cite{roope-intel-fv-i7}, formal methods were deemed so
 successful in examining critical arithmetic units of Intel's core i7 that
 traditional simulation-based testing was largely eliminated.\footnote{In recognition of this success, the leader of
 this project,
 Roope Kaivola, won Microsoft's prestigious verified software award of 2014.}

 Achieving this degree of adoption of formal
 methods in HPC is a coveted goal.
 However, driving a formal methods agenda forward in HPC requires prudence, given the absence of an obvious failure cost model (as happens when chips emerge with silicon defects, where each mask re-spin costs millions of dollars), and
also given the sheer complexity of 
HPC software.
More practical are approaches where 
formal methods are baked into tools so that
everyday users are not confronted with
modeling their idiosyncratic pieces of
code.
\begin{WrapText}
\footnotesize
Correctness of systems hinges
on having validated specifications
and
verification methods that find 
defects.
Challenges in these areas with
respect to
HPC include
the oracle problem, nondeterminism,
performance focus, concurrency,
scale, domain-specific mathematical
abstractions, the use of
floating-point arithmetic, 
and issues that stem from the
underlying programming language
and runtime support.
\end{WrapText}

\subsection{What is the Correctness Problem?}
\label{sec:correctness_general}


A program is \emph{correct} when it behaves as expected on any execution.  This definition begs the question, what behavior is expected?  This is in general a difficult question to answer and will naturally vary from program to program.  Hence the question of correctness involves two related activities: \emph{specification}---the process of rigorously defining what a program is expected to do---and \emph{verification}---the process of establishing that a program complies with its specification, i.e., that it is correct.

\subsubsection{Specification}

\paragraph{Generic vs.\ application-specific properties.}  Certain aspects of the specification of a program come ``for free.''  These include requirements imposed by the programming language used to develop the program.  For example, any correct C program should never attempt to read or write to a memory location beyond the bounds of an object, divide by 0, or dereference a null pointer.  These requirements are specified in the C Standard, and are inherited by any program written in C.

The application program interfaces (APIs) of libraries and other language extensions used by the program may impose additional requirements.  The Message Passing Interface (MPI) standard, for example, requires that all processes belonging to a communicator issue the same sequence of collective calls on that communicator.  The OpenMP Standard forbids data races on shared variables.  As with C, violations of these restrictions lead to undefined behavior, and should never occur in a correct program.

As important as these language-level requirements are, they do not suffice for specifying correct program behavior.  The C program\\
\phantom{xxx}\texttt{int\ main()\ \lb\rb}\\
satisfies all such requirements, but will not correctly compute the solution to a partial differential equation or the effective neutron multiplication factor of a fission reactor.  Clearly, additional techniques must be used to specify \emph{application-specific} properties.

\paragraph{Assertions.}  \emph{Assertions} are a standard way of specifying application\--spe\-ci\-fic properties.   An assertion specifies a boolean expression which is expected to evaluate to \emph{true} whenever control reaches that statement.  Most programming languages support assertions in some way.  In C, for example, \texttt{assert} statements are checked at runtime and a diagnostic message is printed if one fails.  Assertions can also be turned off to save time in production runs, but this limits their ability to establish correctness of HPC applications, since many defects appear only at large scale.

While useful for expressing certain correctness properties, assertions are limited to the primitives available in the programming language and cannot easily express relations across different states.  It is difficult to assert ``forall integers $i$, if $0\leq i<n$ then the value of \texttt{x(i)} when control exited this function is twice the value of \texttt{x(i)} when control entered the function.''

\paragraph{Contracts.}  More sophisticated specification systems such as \emph{contract languages} overcome some of these limitations.  For example, the ANSI C Specification Language (ACSL) is used to specify the behavior of C functions.  The language provides first-order quantifiers (``for all'', ``exists'') and many other primitives beyond those available in C.   ACSL function contracts specify pre-conditions (conditions assumed to hold when control enters the function) and post-conditions (expected to hold when control exits); they also allow one to specify relations between the pre- and post-states.

ACSL contracts are inserted as comments in the code, so they do not impact the usual workflow of compiling and executing the program.  Specialized tools (for performing verification or other tasks) use the contracts in different ways.  The Frama-C platform, for example, can be used to verify ACSL function contracts using deductive (theorem proving) techniques.
%
Contracts may also be added with respect
to collective calls in programming models based on the Bulk
Synchronous Parallel (BSP) Model~\cite{DBLP:conf/vmcai/SiegelZ11}.

\paragraph{Certificates.}  In certification systems, proofs or correctness can be idicated as tactics scripts~\cite{bertot2013interactive} (e.g., written in Coq~\cite{coq}).  In these systems, both the proof and the imperative code that runs can be auto-generated from the tactics; this is how the certified compiler CompCert~\cite{leroy2004compcert} and the Certified Kit Operating System CertiKOS~\cite{certikos} are implemented.

\paragraph{Golden models.}  Finally, sometimes the simplest way to specify an algorithm is to provide an implementation.  This implementation could be a simple, inefficient sequential expression of the algorithm.  It can then be used as a ``golden model'' against which production-quality implementations can be compared.  Methods that can establish the functional equivalence of two programs could then be used to verify the production implementation.



\subsubsection{Verification}

It is well-known that the verification problem is undecidable: there does not exist an algorithm that can always answer correctly the question, \emph{does a program satisfy its specification?}.  But a technique does not have to be perfect to be useful, and over the years, a large number of practical verification approaches have been studied and implemented.  Roughly speaking, we may divide these into two categories.

Tools in the first category attempt to \emph{prove} that a program (with specification) is correct.  If the tool succeeds, the program is guaranteed to be correct.  The tool can fail to find a proof for a number of reasons: the program is incorrect, the resources required (e.g., time or memory) exceed what the user can afford, or the tool is just not capable of finding the proof.  Hence these tools can sometimes show a program is correct, but cannot show a program is incorrect.

Tools in the second category attempt to find defects in programs.  If the tool finds a defect, it has shown that the program is incorrect.  However, such tools may fail to find existing defects---because they are not capable of finding such defects, or cannot do so within reasonable resource limits---and they may report ``false alarms''---possible defects which are not actual defects.  Such tools can show a program is incorrect and provide valuable debugging information, but they cannot show a program is correct.

In reality, this distinction is not black-and-white. Rather, these two categories are two extreme points on a spectrum, with most tools falling somewhere in between.  For example, model checking techniques can be used to prove that a program satisfies specified properties within certain finite bounds (e.g., on the number of processes or inputs sizes) but leave open the possibility that a defect exists outside of those bounds.  Contract-based techniques can show that one function in a program is correct under the assumption that other functions behave correctly.   Other approaches can give probabilistic guarantees.  

In what follows, we outline some of the major currents in software verification research and practice.


\paragraph{Testing.} The most widely-used approach to the correctness problem, testing involves executing the program on a selection of inputs and examining the results.  Testing has become a more rigorous discipline over the last 20 years.  A variety of techniques for selecting test sets satisfying certain criteria (e.g., statement, branch, or path coverage) have been explored. Language-specific properties, assertions, and even contracts can be tested.  The main limitation is that testing cannot establish the program behaves correctly on an input not in the test set.  Other limitations in the HPC context are discussed in Section \ref{sec:correctness_hpc}.

\paragraph{Static analysis.}  These automated techniques attempt to reason about a program without executing it.  Compilers use static analyses to prove properties such as: a variable is never used before it is defined; a variable is only assigned a value of a compatible type; and control never reaches the end of a function body without issuing a \emph{return} statement.  The types of properties that can be proved are generally simple (see Table~\ref{table:tools}).

\paragraph{Dynamic analysis.} In this approach, properties are checked as a program executes, or after the program stops using traces that are gathered when the program executes (see Table~\ref{table:tools}).  Like testing, specific inputs are needed, but dynamic analyses can detect defects that are not normally detected by testing, such as the occurrence of a ``potential deadlock'' even when no actual deadlock occurred during the execution.

\paragraph{Deductive reasoning \cite{hoare:1969:axiomatic}.}  This family uses theorem-proving techniques to prove a program satisfies its specification.  They can be fully automated or require substantial human interaction.  Verification Condition Generation is one increasingly popular approach that generates a number of small theorems from a program+specification which can then be independently ``discharged'' (proved) using a variety of theorem provers.  These approaches often require at least some help from the user, such as code annotations (e.g., loop invariants) or guidance through more difficult proofs.

\paragraph{Symbolic execution \cite{king:1976:symbolic, klee:osdi:2008, siegel-zirkel:2011:tass-mcs}.}  These techniques ``execute'' a program in an abstract sense, using symbols ($X_1$, $X_2$, \ldots) in place of concrete values as inputs.  The ``values'' returned by operations are symbolic expressions (e.g., $X_1-2.7*X_2$).  Symbolic execution can be used to generate high-quality test sets automatically, to find bugs, and even to prove properties (usually with some restrictions such as bounds on input sizes or loop iterations).

\paragraph{Model checking \cite{clarke-grumberg-peled:1999:book}.}  This approach is particularly effective for checking temporal properties of concurrent systems, e.g., ``no process calls function \texttt{f} until every process has exited the ghost-cell exchange.''  It is standard in the hardware industry and is the basis of many software verification techniques for parallel programs.  Typical model checking techniques compute a set of reachable states of a finite transition system.  When applied to software this usually enables exhaustive verification of properties with small bounds on the number of processes and other parameters. Model checking can be combined with symbolic execution to cover a wide range of concurrency behaviors and a wide range of inputs.

\paragraph{Certification.}
In the certification approach, proofs are constructed by the programmer along with the software.  Proof assistants automate aspects of this task to multiply programmer effectiveness in generating code with associated proofs.  Certifying compilers~\cite{leroy2009formal} preserve the proof through code optimization, to produce optimized code along with the compacted proof, in a certificate, of its correctness.  The certificate can be rapidly checked against the binary, e.g., as the program starts, to ensure that the resulting binary code meets the specification. 

\subsection{Challenges in High Performance Computing}
\label{sec:correctness_hpc}


The correctness problem takes on a number of special characteristics in high performance computing.  Here we enumerate some of the most important points.  These points illustrate why specification and verification are particularly needed now in HPC, and identify specific challenges that will need to be overcome.

\paragraph{The oracle problem.}
HPC programs are often attempting to do new science, so the expected results are usually not known.  This makes traditional testing techniques, in which the actual result computed by the program is compared with an expected result, impossible.  (There are often specific cases in which the expected result is known, but these are exceptional.)  Hence HPC requires verification approaches that do not require knowledge of all expected results.  An example would be a tool that proves the functional equivalence of a complex, optimized implementation of some algorithm with a simple, trusted implementation of that algorithm.


\paragraph{Nondeterminism.}
Many HPC programs are nondeterministic.  One source of nondeterminism is concurrency---varying the interleavings of actions from different threads or processes and the computed results may change.  The transition to exascale is expected to lead to even more nondeterminism; hardware components will dynamically adjust their execution rates; software implementations will embrace asynchrony to save time and energy; and linear algebra libraries will increasingly employ randomized algorithm techniques to achieve asymptotic speedups~\cite{spielman2004nearly}.  Testing becomes extremely problematic for nondeterministic systems,  because a correct execution for some input does not even guarantee the program will behave correctly on a second execution with the same input.

There is no one-size-fits-all approach to nondeterminism.   For many programs, the final result is expected to be completely independent of the program's ``internal nondeterminism.''  For others, the final result is expected to vary in expected ways, for example, any difference should result only from the non-associativity of floating-point operations.  In addition, many HPC algorithms, such as Monte Carlo simulations, rely on randomness in an essential way.  For such ``externally nondeterministic'' programs, new specification techniques may be needed, for example, to express correctness in terms of probability distributions.


\paragraph{Performance-focus.}
In traditional software domains, programmers try to express algorithms in the simplest and most natural ways possible.  This makes code easy to understand, maintain, and modify.  In HPC, there is a tension between these goals and the need for good performance.  Simple algorithms that could be expressed in a few lines of code, such as matrix-matrix multiplication, are often re-written using a combination of optimizations, such as loop tiling, loop permutation, and loop unrolling.  The programmer must also introduce explicit parallelism.  Even though such loop optimizations and loop parallelization can be easily performed by a compiler (automatically or interactively), many HPC programmers persist in performing these optimizations manually, introducing the chance for bugs.  The programs are often highly parameterized, and provide multiple implementations of many functions, since different parameters and versions are needed to obtain adequate performance on different platforms.  All of these forces lead to programs that are considerably more complex than they would be if performance were not an overriding goal.  The increased complexity makes defects much more likely and verification even more necessary.

\paragraph{Concurrency.}
HPC programs are parallel programs.  While some of the verification techniques discussed in Section \ref{sec:correctness_general} are applicable to parallel programs, the vast majority of verification work targets sequential programs. For example, the ACSL specification language is very mature and used by a number of tools, but has no support for concurrency.  Furthermore, modern HPC programs are increasingly \emph{hybrid programs} which invoke multiple concurrency models in a layered approach.  These programs are extremely difficult to reason about informally.  Yet even among those verification tools targeting parallel programs, very few can be applied to hybrid programs.  Finally, the use of weak shared memory consistency models---expected to increase in the exascale era, in order to hide memory latencies---adds another layer of complexity and will require new verification techniques.


\paragraph{Scale.}
Modern HPC programs are intended to run at an extreme scale, with astronomical input sizes, numbers of processes or threads, execution time, and so on.  Often, defects are not observed at small scale.  This makes traditional testing and debugging techniques difficult.  It can be very expensive and difficult to obtain time on the machines that can support that scale.  It can take a tremendous amount of time to run tests at that scale.  And debugging a trace involving millions of steps and thousands of threads is an extreme sport.  Traditional model checking techniques also scale poorly.  Therefore HPC requires (1) verification techniques that can scale to that massive scale, (2) techniques that ``downscale'' programs so that defects that normally manifest only at large scale will manifest in the downscaled version, or (3) techniques whose cost is independent of scale.

\paragraph{Mathematical abstractions.}
Many HPC programs use mathematical subjects such as multivariate calculus, differential equations, linear algebra, and (directed) graphs.  Specifying algorithms in these areas is extremely difficult if the specification language does not provide appropriate abstractions, such as \emph{derivative}, \emph{matrix}, or \emph{strongly-connected component}.  Similarly, proof systems or automated verification techniques must be developed to support those abstractions.  Many libraries of this sort exist (see e.g., ~\cite{awesome-coq}) but there is much work to increase their adoption in the HPC community and to fill out needed gaps.

\paragraph{Floating-point arithmetic.}
Many HPC programs involve extensive float\-ing-point computations.  The notion of correctness in such programs is intimately tied up with floating-point issues, such as round-off error.  Increasingly, developers are reducing floating-point precision to reduce communication costs, and the effect of these tweaks on the output is difficult to gauge.  Tools that can analyze the extent of error introduced by these tweaks and determine whether it is within safe margins for a given application are needed.  However, with few exceptions, support for floating-point reasoning is very weak in existing verification systems. Floating-point arithmetic also wreaks havoc on testing-based verification, since it can be difficult to determine the magnitude of an acceptable discrepancy. 

Another aspect of floating-point arithmetic is how compilers treat float\-ing-point optimizations.
All compilers support a slew of ``IEEE-unsafe'' floating-point optimization flags that can yield a manyfold improvement in performance,  but at the expense of changing the results of floating-point calculations. 
The flags themselves vary from platform to platform.
This aspect of floating-point result variability can render applications incorrect, especially if applied with a performance-focus alone (not minding correctness or result-reproducibility).

\paragraph{Programming language.}
Most HPC programs are implemented in Fortran or C++ (or both), while many verification tools target C or Java.  While many of the ideas and even specific techniques are language-independent, significant engineering effort is required to extend existing verification tools to new programming languages.



\section{State of the Art and Successes\ASGNMT{Ignacio (lead), Ganesh, Koushik}}
\label{sec:state-of-the-art-successes}
\newcommand{\subheader}[1]{ \textbf{#1}}


\begin{WrapText}
\footnotesize

 There have been several
 notable successes in establishing
 rigorous methods in support for HPC.
 Many of today's successes lie in
 the use of
 static analysis, dynamic analysis,
 focused testing with non-determinism
 control, anomaly detection specific
 to HPC, and debuggers focused on HPC.
 The use of rigorous and systematic
 methods in many recent projects, while
 not as mature,
 has already shown considerable
 promise.
\end{WrapText}

Today’s correctness practices comprise a body of domain-specific testing,
and tools and frameworks to debug, pinpoint, and fix errors that escaped
the testing phases. Most of these practices are ad hoc---they often require
heavy-weight program instrumentation and analysis, and are tailored to 
specific classes of bugs (e.g., data races), programing models (e.g., MPI),
and runtime systems and platforms. In addition, they are largely not composable,
and are often difficult to adopt in practice in the workflow of large scientific
code bases.
As a result, it is not uncommon for programmers to end up chasing 
elusive bugs by ``printf'' debugging. When an error is reproducible, 
parallel debugging tools can be very helpful in diagnosing an error, 
though this process tends to be manual and requires a significant 
amount of domain expertise.

We split the state-of-the-art practices into two broad categories:
\textit{testing} and \textit{tools for bug detection and localization}.

\newcommand{\CM}{\checkmark}
\definecolor{LightCyan}{rgb}{0.88,1,1}
\definecolor{Gray}{gray}{0.9}

\begin{table}[th!]

\renewcommand{\arraystretch}{1.1}
\centering
\caption{Some of the existing tools and frameworks to detect and localize bugs in HPC programs}
\label{table:tools}
\scalebox{0.80}{
\begin{adjustbox}{width=1.3\textwidth,center}
\begin{tabular}{p{13cm}|lllllll}
 & 
\rot[65]{Formal Method} & 
\rot[65]{Static Analysis} & 
\rot[65]{Dynamic Analysis} & 
\rot[65]{Control of Non-determ.} & 
\rot[65]{Anomaly detection} & 
\rot[65][3em]{Parallel debugging} 
\\ \hline

\rowcolor{Gray}
\multicolumn{7}{c}{\textbf{Serial Code}} \\ \hline
\textbf{Clang Static Analyzer}--{static analysis bug detection in C/C++~\cite{ClangStatic}} & \CM & \CM &  &  &  &  \\ \hline
\textbf{Clang Sanitizers}--{runtime bug detection (e.g., AddressSanitizer)~\cite{CLANGSANITAZERS}} &  & \CM &  &  &  &  \\ \hline
\textbf{Klocwork}--{on-the-fly, scalable static analysis~\cite{KLOCWORK}} & & \CM &  &  &  &  \\ \hline

\rowcolor{Gray}
\multicolumn{7}{c}{\textbf{Multi-threaded Code}} \\ \hline
\textbf{Valgrind}--{memory management error detection and threading bugs~\cite{Valgrind}} &  &  & \CM &  &  &  \\ \hline
\textbf{Intel Inspector}--{memory and threading error debugger~\cite{IntelInspector}} & & \CM & \CM &  &  & \CM  \\ \hline
\textbf{CUDA-MEMCHECK}--{memory access errors detection in GPU code~\cite{CUDAMEMCHECK}} &  &  & \CM &  &  &  \\ \hline
\textbf{ThreadSanitizer}--{data-race detection for multi-threaded programs~\cite{ThreadSanitizer}} &  &  & \CM &  &  &  \\ \hline
\textbf{ARCHER}--{data-race detection for OpenMP programs~\cite{ARCHER}} &  & \CM & \CM &  &  &  \\ \hline
\textbf{GMRace}--{data-race detection in GPU programs~\cite{GMRACE}} &  & \CM & \CM &  &  &  \\ \hline
\textbf{GKLEE}--{concolic verification GPU programs~\cite{GKLEE}} & \CM & \CM & \CM &  &  &  \\ \hline
\textbf{GPUVerify}--{static (SMT-based) verification of GPU programs~\cite{gpuverify:oopsla13}} & \CM & \CM &  &  &  &  \\ \hline
\textbf{DTHREADS}--{deterministic execution of multi-threaded programs~\cite{DTHREADS}} &  &  & \CM & \CM &  &  \\
\hline\textbf{CUDA-GDB}--{NVIDIA CUDA gdb-based debugger} &  &  &  &  &  & \CM \\
\hline\textbf{Insure++}--{runtime error detection~\cite{INSURE++}} &  &  & \CM &  &  &  \\ \hline

\rowcolor{Gray}
\multicolumn{7}{c}{\textbf{Multi-process Code}} \\ \hline
\textbf{MUST}--{MPI deadlock detection~\cite{MUST}} & \CM &  & \CM &  &  &  \\ \hline
\textbf{UMPIRE}--{dynamic error detection for MPI~\cite{UMPIRE}} &  &  & \CM &  &  &  \\ \hline
\textbf{ISP}--{dynamic formal verifier for MPI~\cite{ISP}} & \CM &  & \CM &  &  &  \\ \hline
\textbf{FlowChecker}--{communication errors in MPI~\cite{FLOWCHECKER}} &  &  & \CM &  & \CM &  \\ \hline
\textbf{AutomaDeD}--{anomaly detection in parallel programs~\cite{AUTOMADED}} &  &  & \CM &  & \CM &  \\ \hline
\textbf{Prodometer}--{progress-dependence analysis to diagnose hangs~\cite{PRODOMETER,PACTLaguna:2012}} &  &  & \CM &  & \CM &  \\ \hline
\textbf{Vrisha, WuKong}--{scale-dependent bug detection~\cite{VRISHA,WUKONG}} &  & \CM & \CM &  & \CM &  \\ \hline
\textbf{}{Scale-dependent overflows detection~\cite{INTOVERFLOWS}} &  & \CM & \CM &  &  &  \\ \hline
\textbf{MPIWiz}--{record-and-replay for MPI~\cite{MPIWIZ}} &  &  & \CM & \CM &  &  \\ \hline
\textbf{Retrospect}--{deterministic replay of MPI applications~\cite{RETROSPECT}} &  &  & \CM &  \CM &  & \\ \hline
\textbf{ReMPI}--{record-and-replay for MPI~\cite{REMPI}} &  &  & \CM &  \CM &  &  \\ \hline
\textbf{NINJA}--{noise injection to make ND bugs in MPI manifest faster~\cite{NINJA}} &  &  & \CM &  \CM &  &  \\ \hline
\textbf{SReplay}--{record-and-replay for one-sided communication~\cite{ICS'16,PPoPP'16}} &  &  & \CM &  \CM &  &  \\ \hline

\rowcolor{Gray}
\multicolumn{7}{c}{\textbf{Hybrid (multi-threaded, multi-process) Code}} \\ \hline
\textbf{CIVL}--{formal Verification of parallel programs~\cite{CIVL}} & \CM & \CM & &  &  &  \\ \hline
\textbf{Relative Debugging}--{comparison of two program executions~\cite{DeRose:2015}} &  & \CM & \CM &  &  & \CM \\ \hline
\textbf{STAT}--{stack trace analysis tool~\cite{STAT}} &  & \CM &  &  &  & \CM \\ \hline
\textbf{TotalView}--{parallel debugger~\cite{TotalView}} &  &  & \CM &  &  & \CM \\
\hline\textbf{DDT}--{parallel debugger~\cite{DDT}} &  &  & \CM &  &  & \CM \\
\hline\textbf{LGDB, CCDB}--{Cray command-line parallel and comparative debuggers} &  &  & \CM &  &  & \CM \\

\hline
\end{tabular}
\end{adjustbox}
}
\end{table}

\subsection{Testing}

Although testing scientific software is generally considered to be 
difficult~\cite{kelly2008}, it is nevertheless the mainstay
of today's verification
approaches.
Conventional testing, such as \textit{regression} testing, \textit{white} 
and \textit{black box} testing, and \textit{functional} testing are used to 
check exceptional situations and corner cases. Finer-grained levels of 
testing, such as \textit{unit} testing, are however less common, specially 
in legacy HPC applications~\cite{hovy2016towards}, as the effort of 
generating these tests is difficult to justify for domain scientists.
State-of-the-art testing practices rely on the reproducibility of results 
under fixed inputs, and usually check domain-specific physics laws.
Assertions are used to check expected behaviors and results at different code locations.

Validation through the use of analytical solutions to check results 
against experimental data is also employed to some degree.
Verification is also supported 
through techniques such as 
methods of manufactured solutions
(checking
against solutions to made-up idealized cases)
as well as higher level criteria such
as the order of convergence.


\subheader{Challenges of Testing.}
The main challenge to test scientific codes is the large effort in 
generating test cases, specially for complex multiphysics codes. 
Tests require data input, and in HPC applications this can be very large;
thus exhaustively and manually testing every input is infeasible. 
An option for HPC codes is to scale down the domain, but is often 
infeasible to do without introducing inconsistencies.
HPC codes tend to use user-defined data types and complex and long data
structures, which may be passed through functions, and initializing 
these structures to create different test cases is a huge effort. 
Non-determinism and lack of tool support are other important impediments to testing.

Most testing today is limited to a small-scale setting (small number of processes and threads,
and small input size). HPC resources are shared and it is practically 
impossible (or at least very costly for an HPC center) to perform frequent testing 
(e.g., nightly testing) of all applications at large scale.
This limits the scope of bugs that can be covered by testing---it is expected 
that the behavior that is checked at small scale extrapolates to large scale,
though that is often not the case in practice.

Some of the tools that are used to test HPC software include: tools to write 
regression tests for numerical software, such as ATS (Automated Testing System)~\cite{ATS}
developed at LLNL, continuous integration frameworks, such as Bamboo~\cite{Bamboo},
and C/C++ testing frameworks, such as Google Tests~\cite{GoogleTests}, and 
Boost Tests~\cite{BoostTests}.

\subsection{Infrastructure for Bug Detection and Localization}

There exists a variety of tools and techniques that have been proposed to
detect and to isolate software defects in HPC applications. We categorize
these frameworks in six groups:
\textit{static analysis},
\textit{dynamic analysis},
\textit{formal methods},
\textit{anomaly detection},
\textit{non-determinism control}, and
\textit{parallel debugging}.
We present a short definition of each of these methods as follows, and Table~\ref{table:tools}
lists some of these tools. Note that different methods are not mutually exclusive 
and it is common for tools to use a combination of methods; for example, a tool may 
perform static analysis in one phase, and then to perform dynamic analysis
or formal verification in a another phase.

\subsubsection{Static Analysis}
Static analysis examines the code without executing the program and it is 
perhaps the first line of defense against bugs for programmers. These checks
are typically performed when the program is compiled and can warn the programmer
of possible errors in the program. At the moment of writing, the Clang compiler
has currently more than 670 diagnostic flags. Static checks are performed as well
in Integrated Development Environments (IDE), which can detect errors even before the
compilation phase (Eclipse~\cite{Eclipse}). Klocwork~\cite{KLOCWORK} is an on-the-fly static code analysis
tool that is used at LLNL and other DOE laboratories to detect bugs at early stages. 

More advanced static analysis tools can reason about the semantics of code and find bugs
that traditional compiler warnings cannot find. These tools may use symbolic execution
and abstract interpretation techniques to explore all execution paths in the program.
An example in this category is the Clang Static Analyzer~\cite{ClangStatic}.

While compilers perform a large number of static checks,
this all relies on compilers being correct themselves. However, compilers
can have bugs that often arise when performing optimizations (specially under
concurrency~\cite{chakraborty2016})---these in turn may yield application bugs in 
extreme cases that are very hard to isolate. The test and check of code 
transformations that are semantic preserving
are an active area of research~\cite{leroy2009formal}. Commercial compiler 
vendors dedicate major resources to assembling test cases and 
regression testing and have years of experience in the engineering of 
compilers for correctness and performance; this is why the best 
commercial compilers continue to outpace their open source counterparts
in correctness.

\subsubsection{Dynamic Analysis}
Most of the existing bug detection and localization tools for HPC perform
dynamic analysis~\cite{CACM:Debugging}.
Dynamic analysis involves checking correctness by executing the program with an specific
input (or a set of inputs). There are two broad categories of dynamic analysis, 
\textit{online} and \textit{offline}; in the former, checks are performed during the 
application's execution time, whereas in the latter the checks are performed after
the application has finished execution, usually by analyzing traces of the application
that were gathered during the application run. For HPC programs that run on multiple 
processes (e.g., MPI programs), traces are usually gathered from all processes and
then aggregated for further analysis.

A large group of dynamic checkers are \textit{memory checkers} since many bugs
arise due to incorrect use of memory. The Valgrind memory checker~\cite{Valgrind}, for example,
supports MPI programs and can perform memory checks in all MPI processes. A subgroup
of memory checkers, detects data races in multi-threaded programs, including checking
in heterogeneous systems with accelerators. The Intel Inspector~\cite{IntelInspector} 
and ThreadSanitizer~\cite{ThreadSanitizer} support data-race detection of pthread programs.
ARCHER~\cite{ARCHER} performs data-race detection in OpenMP programs on top of 
ThreadSanitizer and static analysis.

Other dynamic analysis frameworks for bug detection are tools to detect 
deadlocks and synchronization problems in MPI (e.g., MUST~\cite{MUST}, ISP~\cite{ISP}, and DAMPI~\cite{DAMPI}), 
tools to detect errors at the message-passing layer (e.g., FlowChecker \cite{FLOWCHECKER}), and tools to perform progress analysis of processes to isolate the origin of hangs (e.g., Prodometer~\cite{PRODOMETER,PACTLaguna:2012}).

In hybrid programming models, data races occur easily and
are notoriously hard to find.  Conventional state-of-the-art data race
detectors exhibit $10\times-100\times$ performance degradation and do
not handle hybrid parallelism.  UPC-Thrille~\cite{ICS'13,PPoPP'13,SC'11,upc-thrille}
is the first complete
implementation of data race detection for distributed memory
programs. In benchmark programs, UPC-Thrille found all previously known 
data races with at most 50\% overhead when running on 2048
cores.

Finally, dynamic analysis techniques have been proposed to tune the
precision of floating-point programs. Precimonious~\cite{SC'13} is a dynamic
analysis approach that performs a search on the types
of the floating-point program variables trying to lower their
precision subject to accuracy constraints and performance goals.
Blame Analysis~\cite{ICSE'16b} can be used to further speedup the precision tuning of
Precimonious. Blame Analysis functions by executing floating-point
instructions using different levels of accuracy for their
operands. Evaluation on ten scientific programs shows that Blame
Analysis is successful in lowering operand precision.

\subsubsection{Formal Methods}
Formal methods, which allow specification and verification of software, 
haven been used to certain degree in HPC. The SPIN model checker~\cite{SPIN} 
has been used in various approaches to check properties of parallel
programing models, including MPI~\cite{MPISPIN} and 
distributed task-based models~\cite{Murthy:2016}. 
CIVL~\cite{CIVL} is a symbolic execution-based verifier 
that can analyze programs using many HPC-relevant parallel programming models, 
including MPI, OpenMP, Pthreads, and CUDA. 
The ARCHER race detector~\cite{ARCHER} based on formal loop carry independence 
analysis and happens-before analysis detects race conditions in OpenMP programs.
Verification of producer-consumer synchronization achieved through the use of named barriers is studied in~\cite{DBLP:conf/pldi/SharmaBA15}.
Additional success cases of formal methods are listed in Section~\ref{sec:sw-fv-promising-methods}.


\subsubsection{Control of Non-determinism}
When debugging a parallel program, programmers must first reproduce
the bug; however, because of the non-determinism that comes from 
parallelism and non-deterministic inputs, reproducing bugs can be a challenge.
Some data- and message-race bugs, only manifest themselves
one time every many (possible hundreds) runs. Thus, programmers often use tools
to control the non-determinism of parallel programs when debugging.
A common method is to use \textit{record-and-replay} techniques~\cite{REMPI} to record
the execution of a program when the bug manifests, and then to replay the same
execution deterministically using a parallel debugger. Other tools allow
programmers to speedup the manifestation of the bug, i.e., to make it manifest
with more likelihood in less runs (NINJA~\cite{NINJA}).
SReplay~\cite{ICS'16,PPoPP'16} is the first software tool for
deterministic record and replay for one-sided communication. A key
innovation in SReplay is that it allows the user to specify and record
the execution of a set of threads of interest (sub-group), and then
deterministically replays the execution of the sub-group on a local
machine without starting the remaining threads. 

\subsubsection{Anomaly detection}
Anomaly (or outlier) detection---detection of behavior that is significantly
different from the expected (or normal) behavior---can be used to isolate 
software defects. Here, \textit{behavior} can be broadly defined in terms 
of performance or correctness metrics, from slower-than-usual execution 
times to out-of-range floating-point computations or unusual logic actions 
(e.g., some branches taken more often than others). Most methods in this 
domain use traces that are obtained under error-free runs to define 
\textit{normal} behavior, and then traces that are collected when 
an error manifests are used to detect and localize problems.
Some of the work in the area include DMTracker~\cite{DMTRACKER}, Mirgorodskiy et al.~\cite{Mirgorodskiy2006},
AutomaDeD~\cite{AUTOMADED}, and Bronevetsky et al.~\cite{Bronevetsky2012}. 
Anomaly detection has been used as well to detect scale-dependent bugs, 
i.e., bugs that hide themselves at small scale but that manifest at 
large scale (Vrisha~\cite{VRISHA}, WuKong~\cite{VRISHA}).

\subsubsection{Conventional Parallel Debugging}
Parallel debuggers allow programmers to control and to examine the state
of threads and processes in a parallel program. These tools have advanced
graphical interfaces that support a wide range of features to visualize 
the value of variables in the program and can operate under several parallel programming
models, including OpenMP, CUDA, and MPI. Two of the most used commercial options
are TotalView~\cite{TotalView} and DDT~\cite{DDT}.

A very effective way to debug large-scale parallel programs is stack trace analysis; 
the STAT~\cite{STAT} tool provides a lightweight method to gather and merge 
stack traces of parallel processes and to present them to programmers in an 
intuitive way. Relative Debugging~\cite{DeRose:2015} can assists programmers to
locate errors by observing a divergence in relevant data structures between two
versions of the same program as they execute, and is particularly
effective when code is migrated from one platform to another.

LGDB (Cray Line Mode Parallel Debugger) is a GDB-based parallel debugger
developed by Cray that is used in DOE scientific computing facilities, such
as the National Energy Research Scientific Computing Center (NERSC) and
Argonne Leadership Computing Facility’s (ALCF).
CCDB is a GUI tool for comparative debugging that runs LGDB underneath.
Its interface makes it easy for users to interact with LGDB for debugging.

\subsection{Correctness through Correct-by-Construction Certification}
As mentioned above, in the certification approach, a rigorous software development methodology involves writing the proof of code correctness along with the code, using a proof assistant.  In some cases, one does not write the code at all -- it is auto-generated, along with the proof, and the layer specifications, from a sketch written in a tactics language~\cite{WengPhD2016}. Recently this approach has showed great promise in the systems software field, with certifying compilers and ways of developing efficient code along with strong proofs of correctness. The certified software is engineered for proof modularity, so that independently certified parts can be linked for ensuring correctness of the overall system.  The scope of proofs within this community includes reasoning about concurrency, security, storage systems and floating point correctness.  A rich library of code, proof objects, and mathematical ontologies are available for developers to draw on in creating larger systems.  There is great pick-up of this technology within the computer systems research community---it is now expected and rewarded by the top conferences that new systems software technologies are accompanied with the formal proofs of correctness~\cite{chen2015using}.  The certification approach is also rapidly gaining traction in the embedded computing field (for correctness of systems with respect to safety requirements) and cybersecurity field (for proofs of freedom from particular vulnerabilities). Although the software engineered with this approach is highly modularized, e.g., into ``Deep Specifications''~\cite{gu2015deep}, the certification approach does not impose significant performance penalties.  Full and performant concurrent operating systems are available that are fully certified~\cite{gu2016certikos}.  Hardware cores have been designed that export their functional properties (e.g., opcode semantics, memory model, and synchronization semantics) and are formally verified to these specifications.  The specifications are exported so that the software above can be certified in the context of proved hardware semantics.  
\ignore{==>
The National Science Foundation (NSF) has funded a broad multi-institution project ``DeepSpec'' to advance the field of deep specifications.<==}
Recently there has been progress in applying certification approaches to randomize algorithms~\cite{BEGGHS16}, which might lead to ways to certify numerical methods based on statistical assumptions.

\subsection{Successes due to rigorous and systematic methods}
\label{sec:sw-fv-promising-methods}

We enumerate some recent advances in the field of verification that are
related to HPC. While small scale and initial prototypes, and in some 
cases very difficult to achieve, they indicate the feasibility of 
verifying HPC programs, the availability of powerful initial tools, 
and that the field is primed for success.

\begin{itemize}
\item Certification of C programs with floating point: 
Ramananandro et. al.~\cite{ramananandro2016unified} developed a 
tool VCFloat that demonstrated that “floating-point computation can 
be verified in a homogeneous verification setting based on Coq only.” 
Ramananandro used a new formal specification of IEEE 754 Floating point 
called Flocq~\cite{boldo2011flocq}.

\item Verification of a C numerical solver for a wave equation. 
Boldo et. al.~\cite{boldo2013wave} used a tool Frama-C that statically 
analyzes a C program to produce a proof that can be checked by 
a range of different tools, including Coq with Flocq, and also SMT solvers.

\item Rigorous mixed-precision floating-point tuning 
methods, such as in FPTuner~\cite{DBLP:conf/popl/ChiangBBSGR17}, 
promise to lead to optimization methods that can reduce data movement 
and energy consumption, while providing rigorous absolute-error related guarantees.

\item UPC-Thrille~\cite{ICS'13,PPoPP'13,SC'11,upc-thrille} is the first complete
implementation of data race detection for distributed memory
programs.  The implementation tracks local and global memory
references in the program and it uses two techniques to reduce the
overhead: 1) hierarchical function and instruction level sampling; and
2) exploiting the runtime locality specific to Partitioned Global
Address Space applications. 


\item CIVL~\cite{CIVL} is the first symbolic execution-based verifier 
that can analyze programs using many HPC-relevant parallel programming models, 
including MPI, OpenMP, Pthreads, and CUDA. It has also been applied to 
``hybrid'' programs that use more than one programming model. It has found 
bugs in several shorter examples, including race conditions in an OpenMP 
code offered in tutorials.

\item The ARCHER race detector~\cite{ARCHER} based on formal loop carry independence 
analysis and happens-before race checking helped detect a nasty race condition 
(that previously defied debugging for months)  within HYDRA, a large multiphysics 
application developed at LLNL, which is used for simulations at the National 
Ignition Facility (NIF) and other high energy density physics facilities.


\end{itemize}

We list success cases of rigorous and systematic methods that are outside of HPC
but that exemplify how these methods can successfully be applied in other complex
systems:

\begin{itemize}

\item Since 2011, engineers at Amazon Web Services (AWS) have used 
formal specification and model checking to help solve difficult design 
problems in their systems~\cite{CACM2015amazon}. The use of Temporal 
Logic of Actions (TLA+) helped Amazon find several bugs and to improve 
the overall confidence of their systems. This is an excellent example 
of how formal methods have been used in large real-world 
distributed systems.

\item The STACK~\cite{WangZKS13} analysis system uses symbolic execution 
to identify points in the code that can lead to errors because of 
undefined behavior. In particular, it has been very successful in 
identifying checks that could be removed by the compiler because of 
undefined behavior. The tool was used to find over 150 bugs that got 
repaired in open source programs including the Linux kernel 
and the Postgres database system. The program has also been 
used successfully at companies such as Intel~\cite{STACKnews}.

\item Concolic Testing (also known as DART:
Directed Automated Random Testing or Dynamic Symbolic Execution)
alleviated the limitations of classical symbolic execution by
combining concrete execution and symbolic
execution~\cite{PLDI'05,FSE'05}.  Concolic testing~\cite{CACM'13} has been demonstrated as an effective technique for generating
high-coverage test suites and for finding deep errors in complex
software applications. 
The success of concolic testing in scalably and
exhaustively testing real-world software was a major milestone in the
ad hoc world of software testing and has inspired the development of
several industrial and academic automated testing and security tools
such as PEX, SAGE, YOGI, and Vigilante at Microsoft, Apollo at IBM,
ConBol and Jalangi at Samsung, CATG at NTT Laboratories, and SPLAT,
BitBlaze, jFuzz, Oasis, and SmartFuzz in academia. Some of these tools have been successfully applied to discover critical functional abugs and security vulnerabilities in real-world software.  \ggc{For example, SAGE~\cite{DBLP:journals/cacm/GodefroidLM12}} has found many new expensive security bugs in many Windows applications, and is now used daily in various Microsoft groups. SAGE found one-third of all Win7 WEX security bugs.

\item  
\ignore{===>
Benjamin Pierce's group at the University of Pennsylvania, 
working with Chalmers University and the commercial company QuviQ, 
developed a system for formally specifying the behavior of the <===}
\ggc{In recent work,
file sharing and synchronization services
supported by DropBox
was subject to property-based testing~\cite{MysteriesOfDropbox2016}. This shows 
that large scale distributed systems, developed in an ad-hoc way 
without formal methods, can have major holes, even when in massive use,
and that formal methods can be practically used to close these gaps.}

\ignore{===>, a service in active use with 
more than 500 million register users. Pierce and his colleagues tested 
the service's observed behavior as users, against the model. Building 
on observed deviations, 
they were able to show various ``Surprises'' 
where DropBox could briefly delete a created file, permanently recreate 
a deleted file, and
<===}

\end{itemize}

\section{New Research, Impact\ASGNMT{Sriram (lead)}}
\label{sec:new-research-impact}


In this section, we identify extensions to  existing state-of-the-art practices and research needed to make the correctness techniques described so far work in the context of HPC applications.
\ggc{Our description covers static methods~\S\ref{sec:static-methods}, dynamic methods~\S\ref{sec:dynamic-methods}, debugging \S\ref{sec:debugging}, and pragmatic issues~\S\ref{sec:pragmatic-thrusts}.}

\subsection{Static methods\ASGNMT{Armando (lead), Steve, Ignacio, Sriram}}
\label{sec:static-methods}

\newcommand{\skc}[1]{}

Static techniques for program analysis and verification could help increase confidence in HPC results, as well as reduce development time by reducing the effort needed to track down and fix bugs. 
%
\ggc{This potential research agenda can be broadly divided into the following \ggc{thrusts}: runtime
focused thrust~\S\ref{subsec:rt}, 
numerical algorithms
thrust~\S\ref{subsec:numerical},
specification thrust~\S\ref{subsec:spec},
and thrust towards verification of compilers and
libraries~\S\ref{subsec:compilers-libraries}.}

\subsubsection{Runtime system focused thrust}
\label{subsec:rt}

Many recent successes in applying formal reasoning techniques to systems software could be applied directly to the runtime systems underlying a lot of HPC codes. In particular, the MPI and OpenMP runtimes could be good targets for such an effort. 

\begin{WrapText}
\footnotesize
 
 New research in support of
 the correctness challenges in
 HPC is needed in the areas
 of static methods, methods that
have runtime system focus, 
numerical algorithm focus (with
emphasis on floating-point usage),
and focus on verifying compilers
and libraries.
The use of dynamic methods, debugging,
and many pragmatic thrusts (even smart
IDEs) can go a considerable way.
Rigorous methods are essential for
many aspects of concurrency, including
shared memory consistency models.

\end{WrapText}

There is a wide range of approaches that could be applied to these runtimes; at one end, we could attempt to verify these runtimes for the absence of memory errors and race conditions. At the other extreme, recent successes in the development of fully verified system components, from file systems to compilers, suggest that it would be possible to develop fully verified implementations of at least some parts of the MPI or OpenMP runtimes. 

Another potential target for this approach is the compiler itself; there is already significant work on verified compilers, but additional resources could help push this work to focus on verifying the kinds of optimizations that are most relevant to HPC code, such as cache optimizations.

This agenda would have the benefit of immediately eliminating entire classes of bugs from HPC systems without the need for any buy-in from the HPC application development community; application developers would just  swapp one library for another. 

There are a variety of techniques that can find memory errors and data-races in existing software. Applying these techniques to the runtimes that support most HPC applications could yield some insights, helping improve confidence on such systems. 
%
%
A more ambitious goal would be to gradually rewrite major components of the standard HPC runtime using mechanisms more suited for verification. The goal would to move beyond verifying  the absence of memory errors and data races, and towards verifying important functional properties, such as ensuring that no messages will be dropped by the runtime. In addition to the implementation effort, this approach requires new research into the formalisms necessary for verifying such properties.

\subsubsection{Numerical algorithms focused thrust}
\label{subsec:numerical}

The second thrust would involve verifying the numerical computations themselves, under some assumption that the underlying runtime systems are correct. This has the potential for much bigger payoff relative to verifying the runtime system, \ggc{given that most HPC developers run into errors of their own making more often than they run into MPI bugs.} 

On the other hand, this also requires significant new research around two major issues: reasoning about numerical and floating point computation, and making it possible for HPC users to write specifications.

\paragraph{Automated reasoning for reals and floating point}

Numerical computations can be analyzed at two levels; first, they can be analyzed assuming ideal, real-valued computation. In reality, however, computation involves floating-point numbers. In many settings, the assumption of real-valued computation can help uncover outright bugs in the computation, but reasoning about floating points is necessary to reason about issues such as convergence and precision. 

There has been significant recent work on automated reasoning techniques for real-valued computation. For example, the \ggc{dReal~\cite{dReal}} system is able to reason about logical formulas involving transcendental functions such as $sin$, $cos$, $log$ and exponents, as well as formulas involving integrals. 
\ggc{Support for transcendental functions is provided
by recent tools such as FPTaylor~\cite{fm15-fptaylor} that underlies FPTuner~\cite{DBLP:conf/popl/ChiangBBSGR17}, a rigorous floating-point precision tuner.
These tools are yet to be scaled to the sizes often required by DOE physics codes.}

The complexity of reasoning about numerical code depends on the gap between the specification and the implementation. For example, in checking that a tiled implementation of a computation is equivalent to an un-tiled version, the gap between specification and implementation is relatively small. Such a verification problem does not require deep reasoning about the properties of real-valued or floating-point computation, since both versions are performing the same computation, only in different order, so a modern verification tool can abstract away the details of the floating-point computation and focus on verifying that the loop structures are equivalent. This level of reasoning can be achieved with existing technology, and is indeed performed by systems such as \ggc{STNG~\cite{STNG},} which reasons about the equivalence between low-level stencil implementations and their high-level specifications.

At the other extreme, reasoning about the fact that Conjugate Gradient will indeed find the solution of a linear system of equations is extremely challenging, and requires detailed knowledge of linear algebra to be encoded in the verification system. Doing so is likely quite possible and could yield substantial benefits in correctness. 

For floating-point arithmetic and associated error versus performance tradeoff analysis, formal methods can provide safety nets for enabling what practitioners like to do---i.e., push
on performance while skimping on precision.
Formal methods are essential
to define what is safe for the
situation at hand (error containment, ensuring convergence), as floating-point precision tuning 
\ggc{is often not that effective} without modeling the usage context.

\ggc{\paragraph{Unsafe optimizations for floating point:} One aspect of compilation that has received very little attention is how fast the code can be made by pursuing ``unsafe math'' optimizations (upto 5 times faster for some codes, thus a highly tempting option), and yet, these optimizations introduce far more than the normal IEEE round-off error. A recently developed tool FLiT~\cite{FLiT} is able to portray the number of different answers one can obtain even for a single test routine. Unfortunately, the meanings of compiler optimization flags vary across compilers. All this can lead to result variability to an uncalibrated extent, affecting both correctness and reproducibility. This is another aspect of the aforesaid error versus performance trade-off analysis that merits rigorous support.
}

\subsubsection{Specifications Thrust} 
\label{subsec:spec}

One of the major roadblocks in the adoption of verification and formal reasoning technology is the difficulty of writing formal specifications. There are two ways in which other communities have addressed this problem. First is to focus on general properties that every program should satisfy (such as memory safety or race freedom). Tools can be designed to verify such properties without the need for the tool user to have to provide individual specifications for every program. 

Second is to focus on specific domains and write specification languages that are tailored to those domains. 
In the HPC context, there is a recent trend towards domain specific languages that has been fueled by recent successes in generating very high-performance implementations from high-level domain specific notations. Examples of such high-performance systems include TCE~\cite{baumgartner2005synthesis,hirata2003tensor}, Halide~\cite{DBLP:conf/pldi/Ragan-KelleyBAPDA13} and Spiral~\cite{puschel2005spiral}. 

A move towards domain-specific languages can help sidestep the verification problem and make it possible to introduce verification technology without burdening the user. This can be done in a few different ways. First, verification can serve as a form of translation validation. Most production compilers are developed by large organizations and used on millions of programs,\ggc{ so they have no obvious errors.\footnotemark (the CSmith~\cite{regehr-pldi11} work has shown how buggy production compilers can be).}
\footnotetext{\ggc{“There are two ways of constructing a piece of software: One is to make it so simple that there are obviously no errors, and the other is to make it so complicated that there are no obvious errors.” - Tony Hoare}}
Domain-specific languages are less likely to have either of these characteristics, so they are more likely to have bugs. A general verification infrastructure to guarantee that the output of the DSL compiler is consistent with its input could be very useful. Moreover, the DSL compiler could provide a trace of its derivation steps that could significantly simplify the verification task. 
Verification could also help in those (hopefully rare) cases where the output of the DSL compiler needs to be modified for any reason.

\ignore{===>
More ambitiously, a general DSL construction infrastructure could be developed which incorporates verification as an inherent goal in the DSL. Some of the work on the Fiat~\cite{Fiat} system points in this direction. Fiat is a system that allows an expert to define a DSL, together with proof rules for the DSL so that users can get verified implementations directly from high-level programs in the DSL. This technology has only been applied to domains very far from HPC, such as data-processing, but similar technology could have a big impact in HPC. Another system in the same spirit, Bellmania~\cite{Bellmania}, showed the potential for a similar approach to develop a DSL for dynamic programming problems that could generate verified implementations that were parallel and more cache efficient than the standard dynamic programming implementations. 
<===}

Recent work in the context of stencils~\cite{Kamil:2016:VLS:2908080.2908117} has shown that domain-specific compilers can interact with verification in other ways. In the STNG system, a low-level stencil computation is analyzed to extract a high-level specification of the computation in the Halide DSL. This automatic extraction of the specification can make it possible to leverage the power of the high-performance DSL in the context of a low-level legacy implementation. 

\label{DESL}
A DSL can also be embedded in a regular ``host'' programming language; simple C is perfectly able to express loops for linear algebra, stencils, and solvers in a clear and compact ``textbook'' way and high performance codes can be automatically generated from such specifications~\cite{meister-rstream}; no DSL is needed for such domains.  A Domain-Specific Embedded Language (DESL)~\cite{hudak1997domain} has many advantages such as avoiding a ``tower of Babel'' of many different DSLs, clear semantics for linking with other modules, and benefiting from ongoing research and development for the host language.  Investment in optimization and verification of the host language benefits all programs in that language.  

\subsubsection{Verification of compilers and libraries \ASGNMT{Armando (lead), Kouskik, Sriram, Richard}}
\label{subsec:compilers-libraries}

The compiler technology has advanced enough to automate many optimization steps: loop transformations, data layouts, vectorization, etc. HPC applications increasingly rely on optimizing compilers, auto-tuners, and optimized libraries to achieve portable performance. This trend is advantageous from a correctness perspective. Beyond verifying every manual optimization in an application, ensuring correctness of the compilers and libraries can help us ensure more parts of the software toolchain are correct.

Given the shift toward automated data layout
and iteration-space optimizations achieved through
portability layers such as 
RAJA \cite{RAJA-LLNL-TR}
and 
Kokkos \cite{DBLP:journals/jpdc/EdwardsTS14},
the integrity of such ``tall compilation stacks'' 
can become single points of failure
due to bugs they can introduce in
all their generated code. 
Code
generation may also be able to encompass
the generation of complex data structures that are not feasible for humans to originate.
On the flip side, these stacks can also serve
as
{\em single opportune points of intervention}
for maximally impactful uses of formal methods.

\ignore{
Formal methods can provide the underpinnings
for code generation, for example
for different data layouts. The generated
code can provide a consistent representation, as well
as automation of the trade-off space exploration. 
}

Polyhedral optimizations involve the use of well-specified transformations implemented through complex tool chains. Whereas the test suites associated with the tool chains can catch some bugs, they can be sensitive to initialization values used for inputs \cite{DBLP:conf/popl/BaoKPRS16,schordan2014verification}. Verifying the code generated by a polyhedral optimizer, through a combination of verification, exhaustive testing, and certification, is an attractive yet feasible endeavor.

\ggc{One possible route for a verified HPC compiler is to base it on CompCert~\cite{Stewart:2015:CC:2676726.2676985}.}  Most commercial and open source compilers are implemented with traditional non-certified programming techniques, making their verification difficult.  The route would follow the path of engineering the range of optimizations for HPC on an existing certified compiler. Once this is done, domain-specific embedded languages (DESLs, see Section~\ref{DESL}) implemented in the certified host language and compiler would benefit from the certification capabilities.

\subsubsection{Other thrusts \ASGNMT{Ganesh}}

\ggc{Formal methods based on automata-theoretic modeling can be applied to expressing component interfaces in the form 
of interface automata~\cite{DBLP:conf/sigsoft/AlfaroH01},}
or learning the behavior of code that a human expert
does not understand (the latter has been
successfully applied in the Android 
operating system context).

In the area of formal shared-memory consistency models, formal methods are the only satisfactory approach in that while ad hoc testing and manual reasoning may find missed cases, they do not help provide rigorous guarantees that cover {\em all possible executions} allowed by a memory model.

More importantly, formal methods can eminently 
point to formalized testing adaptations, as
in a recent paper~\cite{DBLP:conf/popl/WickersonBSC17}, where formalizing the underlying relations
of memory models in Alloy
allowed the authors to generate tests that
distinguish subtly different 
memory consistency models.

 \subsection{Dynamic methods\ASGNMT{Costin (lead), Koushik, Ignacio}}
 
\label{sec:dynamic-methods}

Static methods are widely  acknowledged for their soundness and precision, but  face challenges when applied to large realistic code  bases.  Code sizes,  layers of abstraction, and  combinations of programming languages (e.g. C++ and Fortran) all pose problems to  static methods. 

In recent  years, dynamic methods have emerged as a practical and  powerful alternative to  static approaches. Dynamic methods  make  inferences based on  observed execution(s) of the program. While  no  guarantees can  be provided for any other unobserved execution, the hope is these inferences  are generic and useful to developers. Tools such as Valgrind,   and Intel ThreadChecker have widespread adoption in the software community and have been shown to be able to handle very complicated codes, such as the Linux kernel. Compared to static approaches, dynamic methods require a  guided process that invloves developer feedback and steering.

Dynamic symbolic execution (or concolic testing)~\cite{CACM'13,PLDI'05,FSE'05,klee:osdi:2008} is a dynamic analysis method where constraint solvers are used to steer the program execution along various feasible execution paths of a program. Though dynamic symbolic execution has been successfully applied to find subtle bugs and security vulnerabilities in sequential software,  little has been done to scale it for parallel and concurrent software~\cite{SAcav06,SAhvc06}.  Research is needed  to combine conventional model-checking and active-testing techniques~\cite{Spldi08,JNPScav09} for concurrent programs with dynamic symbolic execution to make them work for HPC programs.   

\ignore{==> Ganesh fixed this: 
\todo[inline]{Costin: I believe Ignacio added the text directly  below, but I disagree with it. I use in-situ at scale, when I can't dump state out, at small scale we just dump traces as detailed as they need to be. }
\todo[inline]{Ignacio: I didn't add the text below. I agree with Costin that in-situ can be used at large scale.}
<==}

 
Online \ggc{dynamic} analysis methods have the advantage of being 
 deployable in production environments and 
 in conjunction with the actual libraries 
 available on a platform. Therefore, they are
 practical and can provide 
 guarantees pertinent to a particular realization. However, these approaches cannot store or process complete traces and need to minimally perturb the application. Online analysis can benefit from further research into the identification and analysis of relevant interleavings (the partial order) in the presence of multiple concurrency models.
%
%
Offline \ggc{dynamic} trace analysis methods can afford to perform multiple potentially expensive error analysis passes on the traces from large-scale runs. These methods rely on methods to lower the tracing overhead, including the identification and discovery of relevant events to instrument so as to perform the analyses of interest. 

Both online and offline analysis require research to improve their scalability with concurrency and input size. Often, traces contain low-level operations not immediately correlated with the source level. Examples include basic-block level fine-grained control-flow information or load/store information. 
Formal methods can help narrow the gap
between low level traces and human understanding
of the code. These inverse-mapping relations are 
crucial to explain bugs in higher
level terms.
Formal methods can play
a significant role in critical 
design choices such as 
flowing traces into a checker,
shifting between offline and online
analysis, and the use of statistical (sampling)
based approaches to reduce the amount of
tracing done while providing
probabilistic guarantees (e.g.,~\cite{DBLP:conf/asplos/BurckhardtKMN10}).

If support for automated code transformation is desired, research is needed in developer presentation tools that can provide reverse mappings across multiple levels of abstraction. Ideally, the tools could suggest source-level transformations to fix an identified correctness problem. In a large HPC application composed from many libraries, these analyses should be composable and not interfere in terms of correctness or performance. While they can aid in debugging HPC applications, bugs in these analyses can dissuade user adoption. A well-constructed verified toolbox of analysis can complement verified HPC runtimes in ensuring that the bugs identified are indeed from the user's application.

\subsection{Debugging\ASGNMT{Ignacio (lead), Koushik, Costin}}
\label{sec:debugging}


Traditional debugging tools and techniques help to identify the root cause of errors by allowing programmers to control the application and to inspect the application’s state (e.g., value of variables) in an execution. Parallel debuggers control and inspect the execution of many threads and processes, a task that can be computationally expensive given the high degree of parallelism in today’s largest HPC systems. A disadvantage of these methods is that they are manual in nature, i.e., the programmer has to reason about the program and manually find the bug. Advanced debugging techniques and tools help programmers to automatically pinpoint bugs---some with fine granularity, e.g., lines of code. These automatic methods, however, are mostly dynamic (i.e., they can only make decisions based on a given input and execution) and may suffer from high false-positive rates. There exists complementary techniques that aid in the debugging process, such as record-and-replay techniques, which allow programmers to deterministically reproduce bugs. These techniques are of great help to isolate software defects that manifest themselves rarely or non-deterministically. 

Extensions to the state-of-the-art debugging methods are required in the following areas:

\begin{itemize}
\item  Scalable debugging tools to isolate software defects that manifest at large scale, where scale represents number of threads, number of processes, and/or input size. Two categories are important in this area: (i) scalable tools to help control and analyze a program in a large-scale execution when a bug manifests itself, and (ii) debugging tools to isolate scale-dependent bugs using small-scale runs.

\item Accurate automatic debugging techniques to help programmers automatically find the origin of errors to a fine degree of granularity, such as the line of code, function, or code component. In particular, research is needed to improve the accuracy of existing techniques in this category. Metamorphic testing is promising in this regard~\cite{uplee-kanewala}.

\item Methods to control non-determinism when debugging, such as record-and-replay, thread/process schedule controlling, and thread/process schedule enforcing techniques, are needed.

\end{itemize}


\subsection{Pragmatic thrusts\ASGNMT{Koushik (lead), Steve, Costin, Paul}}

\label{sec:pragmatic-thrusts}

      \paragraph{Smart IDEs.}  In the recent years, Integrated Development Environments (IDEs) have gotten smarter in dicovering bugs and common programming mistakes at development time.  As a programmer types her/his program, these IDEs perform on-the-fly code analysis and instantaneously report syntax errors and complex static errors.  Examples of such smart IDEs include \ggc{Eclipse, Intellij IDEA~\cite{IntelliJ-IDEA}, and CLion~\cite{CLion}.} These IDEs not only perform on-the-fly analysis and report static programming errors, they also utilize state-of-the-art program analysis techniques to help programmers with code refactoring and navigation.  In practice, smart IDEs have been found to significantly improve programmer productivity.  
%
%
In supporting HPC software development, smart IDEs can be extended to find concurrency related bugs, such as data races, deadlocks, and atomicity violations. Existing smart IDEs cannot reason about hybrid programming models often used in HPC programs.  Correctness tools and techniques can be made easily accessible to HPC programmers if the formal program analysis techniques developed for HPC programs can be integrated into these IDEs.  

\paragraph{Software design, specification, and testing practices.}
%
%
An HPC correctness campaign can target a few key steps in the software development lifecycle to improve our confidence in their correctness. First, many of the target DOE applications for correctness verification are monolithic and lack formal specification.  Research is needed into methods for ``reverse engineering'' specifications, such as the lifting technique implemented in \ggc{ Helium~\cite{DBLP:conf/pldi/MendisBWKRPZA15}.}  This process will be helped by the design of tools and techniques to decomposing monolithic applications into verifiable units and composing the results of verification.
Second, conventional software engineering teams employ code guidelines, such as Code C++ guidelines, Google style guide, etc. to avoid common design and programming errors. Many of these coding guidelines are associated with tools that can check for conformance. The availability of such tools for HPC software (e.g., precluding the use of \textsf{MPI\_{}COMM\_{}WORLD} would help improve software quality and end-user's trust in their correctness.
Third, a significant challenge in regression testing of computational science applications is assessing when a change in the program output is significant.  Often, mandating that the output remain bitwise equivalent is too strong a requirement and may not be possible in the case of non-deterministic applications, but selecting an arbitrary numerical tolerance may result in missed bugs.  Research is needed to adapt regression testing to applications with large amounts of floating-point arithmetic.

\paragraph{Bug Repositories.}

Many open-source projects maintain public bug tracking systems, which can be used to identify bugs found ``in the wild.'' These  repositories encourage the development of practically useful tools and to evaluate research tools on real-world bugs. While many HPC projects are open source, the use of bug-tracking systems needs to be promoted among the DOE application developers. Going beyond bug-tracking systems, developing guidelines for sharing bugs and code snippets to reproduce them can accelerate the development of tools that can handle HPC-specific correctness challenges. Ideally, tools can help automatically mine bug repositories to isolate bugs from other sources of errors (software configuration, user errors, etc.), validate the bugs, and extract relevant code harness from the patches used to close a bug.
\ggc{Even the study of job failures on HPC clusters and the reasons for 
such failures  (e.g., \cite{job-failures-in-hpc}) would be valuable for the community to
compile and share.}

\section{Next steps\ASGNMT{Richard (Lead), Sriram, Steve, Koushik, Paul, Ganesh}}
\label{sec:specific-recommendations}

\subsection{ASCR Focus Areas}

There are many opportunities for return on investment in HPC correctness.  These returns and investments would fall into the short, medium and long term time scales.

\subsubsection{Short Term}

\paragraph{Advances for production use.}
Investment focused on supporting the current efforts to program and perform computational science on leadership computing facilities.  Such effort would apply best-of-breed existing tools (commercial as well as those being researched by the community), extend those tools, and generally work with HPC applications code as-is.  Such work would encompass extending the existing HPC tools infrastructure (debuggers, compilers, etc.) with features for larger scale verification and debugging in the more complex HPC contexts being encountered today.

\begin{WrapText}
\footnotesize
 
 We propose many directions
 broken down into short, medium and
 long-term components.
 We also propose
 an agenda for a correctness
 workshop as well as 
 a few   ``moonshot'' projects
 that can bring in added creativity through added time and resource pressures.
 
\end{WrapText}
\paragraph{Importing successful ideas from non-HPC domains.}
This effort would focus on bringing the tools and technologies currently being developed to prove correctness and safety properties of non-HPC code, to the HPC community.  This would bring software currently being used to formally prove hardware, security, safety, and performance---as applicable---to HPC.  Research in the formal verification of cyber-physical systems could be applied to the verification of simulated physical systems.  Research in embedded computing - e.g. verification of controls, and verified hardware, e.g. flight safety, could be applied to verifying the controls for current HPC systems.  Certification techniques for large scale distributed systems (e.g., Amazon, DropBox) could be brought to HPC for reasoning about large scale parallel computing systems (and HPC file systems).

\subsubsection{Medium Term}

\paragraph{Community-wide impact:} Medium term focus must target achieving community-wide impact  through projects that take on the challenges of 
verifying critical software components such as  widely-used runtime systems (e.g., OpenMP and MPI implementations) and math libraries.
There must also be significant emphasis placed on the collection of bug incidence reports, as well as search for past solutions that detail how these bugs were fixed.
Finally,  tool interfaces and runtime event collection mechanisms must be standardized to support tool composition.

\begin{figure}
\label{figure:goals}
\centering
\begin{adjustbox}{width=1.15\textwidth,center}
\includegraphics[width=0.99\textwidth]{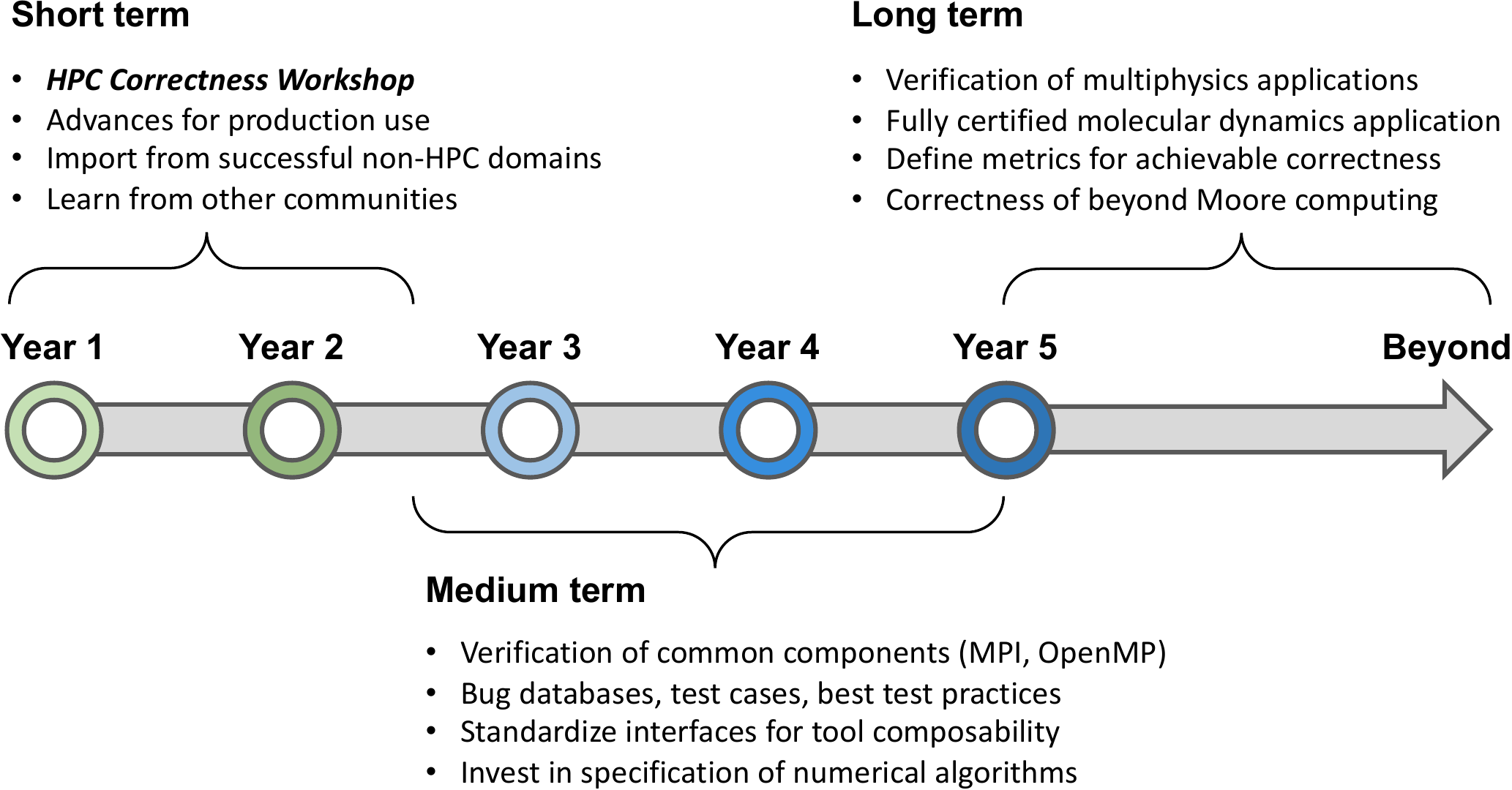}
\end{adjustbox}
\caption{Short, medium, and long term goals identified to advance the field of correctness in HPC}
\end{figure}

\paragraph{Ontologies of mathematics / algorithms.}
This effort would focus on developing specific infrastructure for verification and debugging unique to HPC.  In particular, investments in the modeling and specification of numerical algorithms, ontologies for the mathematics underlying such algorithms, and reasoning about statistical and randomized systems, would be advanced.  There would likely be a close relationship between these efforts and Uncertainty Quantification and Automatic Differentiation, since such efforts are aimed at giving confidence in the results of scientific simulations.  Research would be aimed at tools for automated reasoning and verification about topics in Applied Mathematics of relevance to ASCR, in particular numerical methods for scalable solving of PDEs, discretization, optimization, multi-scale computing, and multi-physics. The effort would be directed toward formalizing mathematics otherwise expressed in ``prose'' in mathematical research papers.  The effort would try to parallel the development seen in the computer systems research community, where new systems software research papers (operating systems, file systems, etc.) are now expected to come with the formal expression of the algorithm and the proof of its correctness.  Such an effort would nudge and enable the engineering of these new mathematical advances to be done with the most modern software engineering practices, with the modularity, specifications, and proofs needed to achieve correct by construction HPC systems.  

\subsubsection{Long Term}
 
\paragraph{Beyond Moore.}
Focus on Beyond Moore computing would clearly be an appropriate  long term focus.  In particular, ensuring correctness of new computing paradigms such as neuromorphic, probabilistic, and quantum computation.  Application areas such as machine learning would also be a focus of the correctness research.  Machine learning in particular seems to offer opportunities for new advances in correctness, particularly developing systems for proving that a ML system will work within bounds, and for explaining the conclusions or decision made by a machine learning system.

\paragraph{``Moonshot'' projects.}
Longer term investments would be achieved by moonshot projects that seek to build end-to-end, demonstrable successes, with both immediate benefits and which also become the basis for advances in the field.
A few potential moonshot projects are described in \S\ref{sec:iafre}.

\subsection{Moonshots}
\label{sec:iafre}

Well-chosen ``moonshot'' challenges can help increase the pace of progress and demonstrate what is possible.  While these projects are ambitious, they are probably feasible with just a few years of focused effort.  Consider that the first sequencing of a human genome took about ten years and billions of dollars, but now such sequencing is routinely practiced within a few hours in a doctor's office lab for a few hundred dollars.  Once feasibility is established in the projects below, the engineering of tools to reduce costs and speed the results will rapidly advance.  For verification in particular, consider that proof libraries and tools are cumulative, and can lead to building of capabilities.  The first efforts to formalize floating point took years to achieve, but the formal specifications for floating point are now available and can 
be downloaded for free.

\subsubsection{Project 1: Fully certified molecular dynamics simulation}  



Molecular dynamics (MD) packages such as Desmond~\cite{bowers2006scalable} would serve as excellent ``moonshot'' projects.  The tools and science of projects such as Deep Specification~\cite{Pierce:SPLASHTalk2016} could be applied, extending them to the special tuned number representations and operators of DESMOND, through bond models and approximations, and through the compiler, systems software, and runtime.  One would certify the property of bit-reversibility through to the implementation.  The components of such a project (e.g., a fully certified concurrent dynamic runtime task scheduler, certified parallel 3D FFT) could then be used in other projects. 

Longer term, this technique could be used to provide certifications and verification that a long running numerical simulation running on a special purpose scientific computer {\em proves} a scientific result~\cite{sussman1992chaotic}.  This would not just be an academic exercise; vital missions of the Department of Energy (e.g., the NNSA) depend on simulation on complex, custom constructed high performance computers to assure the safety and performance of our nation's strategic nuclear arsenal. Advances in NNSA simulations of kinetic plasma on the special Roadrunner supercomputer~\cite{bowers2009advances} used optimizations and coding techniques closely relate to those of DESMOND.  The results of such moonshot projects could immediately carry over to kinetic plasma simulations of the type that runs on Roadrunner.  Furthermore, such a certification system could be used to assess the implications of new hardware architectures (e.g., network communications protocols), representations (reducing the precision of values) and algorithms (e.g., communication avoiding or sparse high dimensional FFT) on future high performance kinetic plasma simulations.

\subsubsection{Project 2: Multiphysics}
Multiphysics software systems---simulations that consists of more than
one component governed by its own principles---are used in many large-scale physical simulations.
An impactful project would be full verification of important logic and numerical properties (e.g., energy conservation properties or others) underlying multiphysics applications.
This exercise can force the examination of how individual subsystem guarantees help meet whole-system correctness goals.
Some examples of HPC multiphysics software infrastructures that could be targeted are Chombo~\cite{Chombo}, PETSc~\cite{PETSc}, SUNDIALS~\cite{SUNDIALS}, Trilinos~\cite{Trilinos}, and Uintah~\cite{Uintah}.

\subsubsection{Project 3: Verified Compiler/Runtime Components}
Verifying some of the key software infrastructural components of an HPC system can bootstrap the development of rigorous methods that help harden support for large-scale runs of HPC simulations.
Of special interest would be the verification of the MPI library, going by the MPI 4.0 standard, tracing through various device layers and ending in optimized infrastructural code that supports rapid messaging using lock-free programming methods.
Similarly, verifying the polyhedral compilation toolchain and linear algebra libraries can ensure correctness of large and highly reused code bases.

\subsection{HPC correctness workshop}
The exercise of bringing this limited set of report authors together for sharing ideas has resulted in good cross fertilization of ideas for HPC correctness: making us aware of useful tools for our own research in HPC, and some of the larger challenges.  But with the breadth of the problem, and the richness of the verification and debugging community outside, HPC, more is needed.

Advancement in this area of correctness could be facilitated by a workshop on HPC correctness that could bring together the larger community of experts on correctness techniques and tools with DOE stakeholders, especially the developers of DOE HPC applications. 

Such a workshop would provide an opportunity for HPC software developers to communicate current practices and discuss the primary obstacles to achieving correctness in HPC software development and for correctness experts to identify promising research directions that offer the potential to overcoming these obstacles.  A two day workshop comprising a small number of invited presentations, 5--10 minute presentations based on 2-page position papers solicited from the community, and 3--4 1-hour round table discussions to stimulate a dialogue between the correctness experts and stakeholders is recommended.  We anticipate that this dialogue will reinforce many of the summit findings, possibly identify additional research opportunities, and help prioritize future research directions.

\subsection{Competitions for verification of HPC software}

In recent years, several verification competitions have evolved within the general software verification community.
For example, the annual SV-COMP competition, currently in its seventh year~\cite{sv-comp}, is a fully automated competition in which participants submit tools which are all fed a long series of C programs with corresponding properties that are expected to hold or fail.  The VerifyThis competition~\cite{verify-this}, in its sixth year, is an interactive competition in which participants are given a set of problems which they are expected to implement and verify over the course of a day using any tools they desire.
These competitions have had several beneficial impacts: they provide objective and consistent comparisons of tools, they provide a recognized measure of the current state-of-the-art, and they have created large verification benchmark suites that are widely used beyond the competition itself.

As discussed above, HPC software has many specific characteristics, and these are not covered in the existing general-purpose software competitions.  Therefore, a HPC-specific verification competition could be held, say, during the course of one day at SC17 with multi-agency sponsorship.   Similar to VerifyThis, participants could be given a set of programs of increasing complexity, together with written specifications of expected behavior, and asked to formally specify and verify as much as they can, using any tools they desire.   A panel of judges would examine and evaluate the results.  Participants could also be given an opportunity to present their solutions.

\clearpage

\bibliographystyle{acm}
\bibliography{HPC-Correctness}

\ignore{=====>
6. Sample Two-Day Correctness Workshop

   a. Themes

   b. Sessions

      1. Keynotes

      2. Roundtables

   c. Outcomes
<=======}

\end{document}